\documentclass[a4paper]{PoS}

\usepackage[utf8]{inputenc}

\title{Leptonic decay-constant ratio $f_K/f_\pi$\\ from clover-improved $N_f=2+1$ QCD}

\ShortTitle{$f_K/f_\pi$ from clover-improved $N_f=2+1$ QCD}

\author{\speaker{Enno~E.~Scholz}\\
        Universit\"at Regensburg, Institut f\"ur Theoretische Physik, D-93040 Regensburg, Germany\\
        E-mail: \email{enno.scholz@physik.uni-regensburg.de}}

\author{Stephan D\"urr\\
        Bergische Universit\"at Wuppertal, Gau{\ss}str.~20, D-42119 Wuppertal, Germany\\ J\"ulich Supercomputing Centre, Forschungszentrum J\"ulich, D-52425 J\"ulich, Germany\\
        E-mail: \email{durr\,(AT)\,itp.unibe.ch}}

\abstract{%
The leptonic decay-constant ratio $f_K/f_\pi$ is calculated from lattice-QCD simulations using $N_f=2+1$ dynamical fermion flavors in the clover-improved formulation and 2-HEX smearing. The simulations were performed for a range of mass-degenerate light quarks including the physical point and at various lattice couplings and volumes, allowing to quantify all relevant sources of systematic uncertainties for our final number of the decay-constant ratio. Utilizing input from chiral perturbation theory, we also quote the charged decay-constant ratio $f_{K^\pm}/f_{\pi^\pm}$. With further input from super-allowed nuclear $\beta$-decays, eventually we obtain an estimate for the CKM-matrix element $V_{\rm us}$.%
}

\FullConference{34th annual International Symposium on Lattice Field Theory\\
		24-30 July 2016\\
		University of Southampton, UK}

\newlength{\closercaption}
\newcommand{\cCap}{\vspace*{\closercaption}}

\newlength{\afterTable}
\newlength{\afterFigure}
\newcommand{\aFig}{\vspace*{\afterFigure}}

\newlength{\closersection} 
\newcommand{\cSec}{\vspace*{\closersection}}

\begin{document}

\section{Introduction}
\label{sec:intro}
\cSec

In order to determine the ratio $V_{\rm us}/V_{\rm ud}$ of Cabibbo-Kobayashi-Maskawa (CKM) matrix elements\footnote{Here we adopt the convention, that the CKM-matrix elements $V_{\rm ud}$ and $V_{\rm us}$ are real and the complex phase in the first row only shows up in $V_{\rm ub}$.} from experimentally measured decay widths of the leptonic kaon and pion decays according to \cite{Marciano:2004uf}
\begin{equation}
\label{eq:marciano}
\frac{\Gamma(K^\pm\to\ell\nu_\ell)}{\Gamma(\pi^\pm\to\ell\nu_\ell)}\;=\;\frac{V_{\rm us}^2}{V_{\rm ud}^2}\:\frac{f_{K^\pm}^2}{f_{\pi^\pm}^2}\:\frac{M_{K^\pm}^2}{M_{\pi^\pm}^2}\:\frac{\left(1-m_\ell^2/M_{K^\pm}^2\right)^2}{\left(1-m_\ell^2/M_{\pi^\pm}^2\right)^2}\:\Big(1+\delta_{\rm em}\Big)
\end{equation}
one needs to know---besides the radiative correction $\delta_{\rm em}$---the ratio of the decay constants $f_{K^\pm}/f_{\pi^\pm}$, i.e.\ the ratio of the hadronic form factors for these decay channels. Here we will report on a calculation of this decay-constant ratio obtained in the isospin symmetric limit of three flavor QCD ($m_{\rm up}=m_{\rm down}\neq m_{\rm strange}$ or short $N_f=2+1$ QCD) without taking into account QED effects. More details of this analysis already appeared in \cite{Durr:2016ulb}.

The report is organized in the following way: in the next section we will present the data used in this analysis as well as the methods to obtain the decay-constant ratio $f_K/f_\pi$ in the isospin symmetric limit without electromagnetic corrections. In Sec.~\ref{sec:disc} we then will convert our result to one for $f_{K^\pm}/f_{\pi^\pm}$ and discuss the implications for the CKM-matrix elements $V_{\rm ud}$ and $V_{\rm us}$.

\section{Decay-constant ratio at the physical point}
\label{sec:physPoint}
\cSec

In this section we will first introduce the ensembles used for our analysis of the leptonic decay-constant ratio. Next we will discuss the functional forms used for the necessary inter- and extrapolations to physical quark masses and the continuum as well as the infinite-volume limits, respectively. Also the fit ranges applied to our data will be discussed before, eventually, our result for $f_K/f_\pi$ is presented with all sources of uncertainties considered.

\subsection{Ensembles used in the analysis}
\label{subsec:ens}
\cSec

\begin{figure}[t!]
\begin{center}
\includegraphics[width=.45\textwidth]{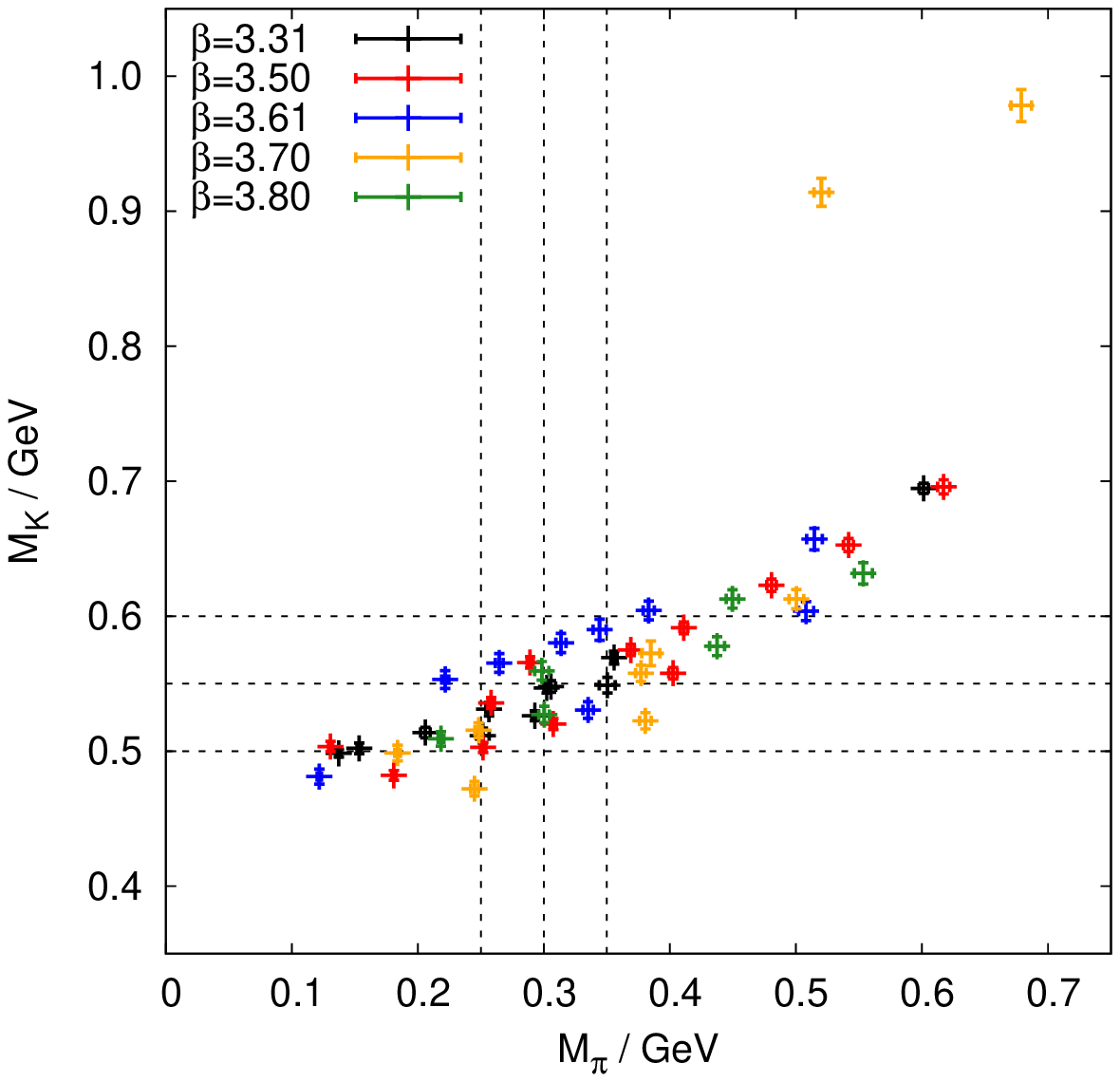}%
\hspace{.05\textwidth}%
\includegraphics[width=.45\textwidth]{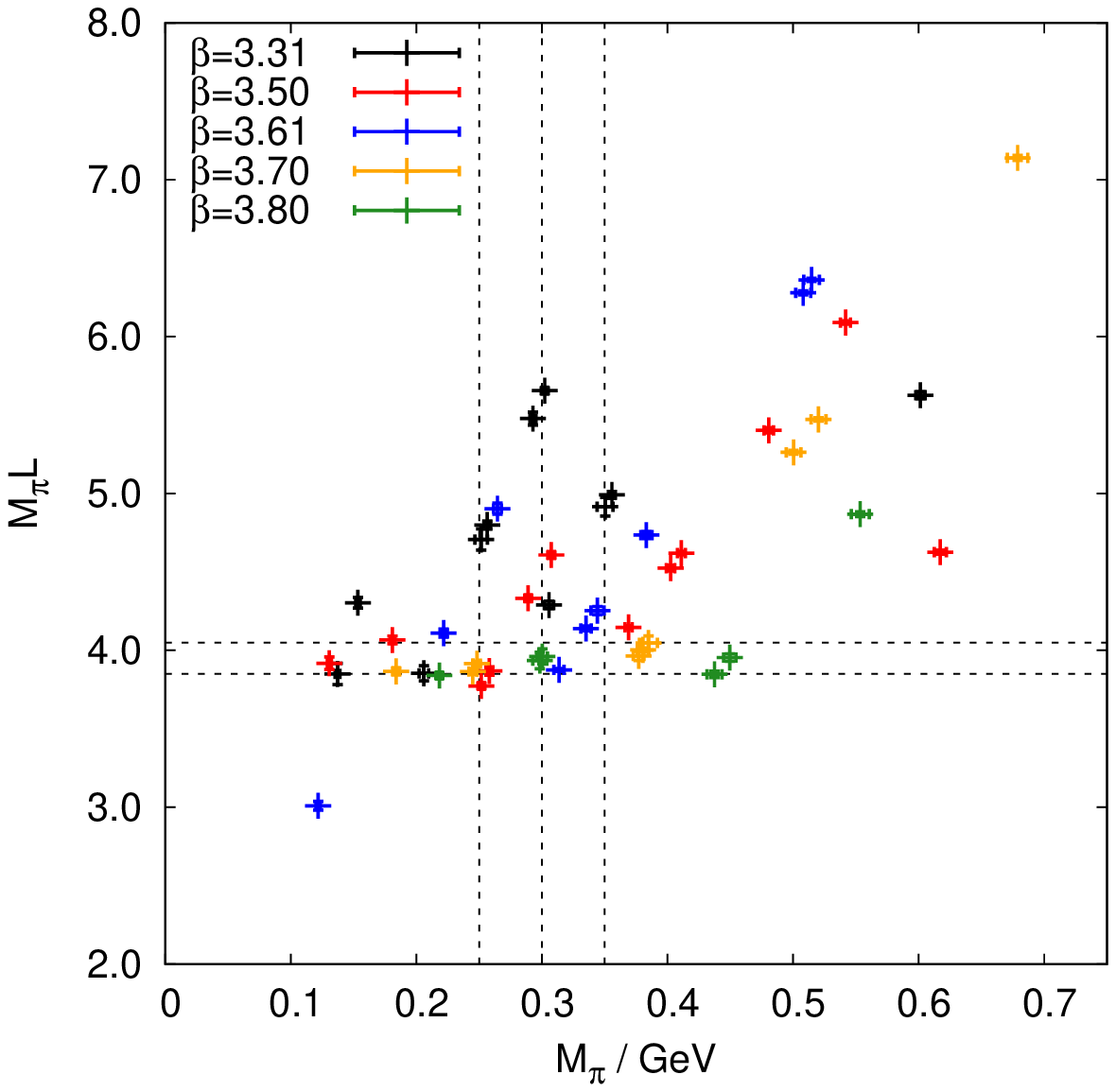}
\cCap
\caption{Overview of the ensembles used in this analysis. The {\it left panel} shows the kaon mass $M_K$ as function of the pion mass $M_\pi$. To convert from lattice units to GeV, the mass of the $\Omega$-baryon has been used to determine the lattice scale $1/a$ for each set of ensembles at the same gauge coupling $\beta$ (mass-independent scale-setting). The {\it right panel} shows $(M_\pi L)$ as function of $M_\pi$. The {\it vertical} and {\it horizontal dashed} lines indicate the fit ranges used as discussed in Sec.~\protect\ref{subsec:fits}}
\label{fig:ens}
\end{center}
\end{figure}

In total, 47 ensembles of gauge configurations generated by the Budapest-Marseille-Wuppertal collaboration (BMW-c.) with $N_f=2+1$ flavors of tree-level clover-improved Wilson fermions and the tree-level Symanzik-improved gauge action were used. See \cite{Durr:2010vn,Durr:2010aw} for more details on the generation of these ensembles and other analyses carried out using them. The ensembles were generated at five different values for the gauge coupling $\beta$, resulting in lattice scales $1/a$ in the range from 1.7 to $3.7\,{\rm GeV}$ (using the $\Omega$-baryon mass to set the scale). The (degenerate) light quark masses $m_{\rm ud}$ were such that pion masses between 130 and $680\,{\rm MeV}$ were obtained. In all but two cases the strange-quark mass was chosen close to its physical value. In Fig.~\ref{fig:ens} we show the kaon mass $M_K$ as a function the pion mass $M_\pi$ (left panel) as well as the product $(M_\pi L)$ as a function of $M_\pi$ for the 47 ensembles used in our analysis. Here $L$ is the spatial extent of the simulated lattice. More details and tables containing simulation parameters and measured quantities can be found in \cite{Durr:2016ulb}. For details on how the decay constants of the pion and the kaon, $f_\pi$ and $f_K$, resp., were obtained, we refer to \cite{Durr:2013goa}.

\aFig

\subsection{Functional fit forms and fit ranges}
\label{subsec:fits}
\cSec

We have to interpolate our data to the physical quark masses as well as to extrapolate to the continuum and infinite-volume limits. In order to be able to quantify the systematic uncertainty associated with each of these inter- and extrapolations, we will always consider at least two different choices for the functional forms used and apply various fit ranges. Finally, we will take the variation of the results from the various fits as our systematic uncertainty.

For the interpolation to the physical mass point, given by $M_\pi^{\rm phys}=134.8(0.3)\,{\rm MeV}$ and $M_K^{\rm phys}=494.2(0.4)\,{\rm MeV}$ following the recommendation of the FLAG report \cite{Aoki:2013ldr}\footnote{At the time of writing the original article \cite{Durr:2016ulb} the latest FLAG report \cite{Aoki:2016frl} was not yet available to us. But except for a marginally change in $M_K^{\rm phys}=494.2(0.3)\,{\rm MeV}$ the values used in this work for $M_\pi^{\rm phys}$ and $M_K^{\rm phys}$ are not affected.} which corrects the experimentally measured values $M_{\pi^\pm}$ and $M_{K^\pm}$ for strong isospin breaking and electro\-magnetic effects, either we use a fit form based on SU(3) chiral perturbation theory (ChPT) \cite{Gasser:1984gg}
\begin{equation}
\frac{f_K}{f_\pi}=
1+\frac{c_0}{2}
\left\{
\frac{5}{4}M_\pi^2\log\left(\frac{M_\pi^2}{\mu^2}\right)-
\frac{1}{2}M_K^2\log\left(\frac{M_K^2}{\mu^2}\right)-
\frac{3}{4}M_\eta^2\log\left(\frac{M_\eta^2}{\mu^2}\right)+
c_1[M_K^2-M_\pi^2]
\right\}
\end{equation}
with two fit parameters $c_0$, $c_1$ and $M_\eta^2\;=\;\frac13(4M_K^2\,-\,M_\pi^2)$ and $\mu$ being an arbitrary scale for the renormalization of ChPT.  Alternatively, one of the following polynomial forms with either three, four, or six fit parameters $c_i^{n-{\rm par.}}$ are used.
\begin{eqnarray}
\frac{f_K}{f_\pi} &=& 1\:+\:[M_K^2-M_\pi^2]\Big(c_0^{\rm 3-par}\,+\,c_1^{\rm 3-par} [M_K^2-M_\pi^2]\,+\,c_2^{\rm 3-par} M_\pi^2\Big)
\\
\frac{f_K}{f_\pi} &=& 1\:+\:[M_K^2-M_\pi^2]\Big(c_0^{\rm 4-par}\,+\,c_1^{\rm 4-par} [M_K^2-M_\pi^2]\,+\,c_2^{\rm 4-par} M_\pi^2\,+\,c_3^{\rm 4-par} M_\pi^4\Big)
\\
\frac{f_K}{f_\pi} &=& 1\:+\: [M_K^2-M_\pi^2]\Big(c_0^{\rm 6-par}\,+\,c_1^{\rm 6-par} [M_K^2-M_\pi^2]\,+\,c_2^{\rm 6-par} M_\pi^2\,+\,c_3^{\rm 6-par} M_\pi^4 \nonumber\\
 && \phantom{1\:+\:[M_K^2-M_\pi^2]}\,+\,c_4^{\rm 6-par} M_\pi^2 [M_K^2-M_\pi^2]\,+\,c_5^{\rm 6-par} [M_K^2-M_\pi^2]^2\Big)
\end{eqnarray}
Note that all these fit forms obey the flavor-symmetry constraint $\left.f_K/f_\pi\right|_{m_{\rm ud}=m_{\rm s}}=1$.

For the continuum limit extrapolation, we either assume scaling by $a^2$ or $\alpha a$, where $\alpha$ is the strong coupling parameter at the simulated gauge coupling $\beta$. In each ansatz, one fit parameter appears, see \cite{Durr:2016ulb} for the explicit expressions we applied. In order to also take into account the systematic uncertainty from setting the lattice scale, we adopted two different ans\"atze using the mass of the $\Omega$-baryon (taken from \cite{Agashe:2014kda}): a mass-independent one and an ansatz, were the scale was set per-ensemble, again for details we refer to \cite{Durr:2016ulb} where also the values used for $\alpha$ are tabulated. The two extrapolation ans\"atze for the infinite-volume limit are based on the ChPT-formulae for the volume dependence \cite{Luscher:1985dn,Gasser:1986vb,Colangelo:2003hf,Colangelo:2005gd}, where in each ansatz again one fit parameters appears, see \cite{Durr:2016ulb}. These different choices already result in 32 combinations of fit forms with four, five, six, or eight fit parameters.

The fit ranges in the masses, lattice volumes, and lattice couplings were chosen as follows:
\begin{eqnarray*}
M_\pi &\leq& 250\,{\rm MeV},\:300\,{\rm MeV},\:350\,{\rm MeV},\:\textrm{no bound}\,,\\
M_K &\leq& 500\,{\rm MeV},\:550\,{\rm MeV},\:600\,{\rm MeV},\:\textrm{no bound}\,,\\
(M_\pi L) &\geq& \textrm{no bound,}\,3.85,\,4.05\,,\\
\beta &\geq& 3.31,\,3.50,\,3.61\,.
\end{eqnarray*} 
These are also indicated by the horizontal and vertical dashed lines in Fig.~\ref{fig:ens}. Together with the 32 considered combinations of fit forms and the requirement to only consider ``true fits'', i.e. only fits with at least one degree of freedom, our data allowed for a total of 1368 different fits. For examples of specific fits, again, we refer to \cite{Durr:2016ulb}. The combined results of these fits will be discussed in the following section.

\subsection{Results}
\label{subsec:results}
\cSec

\begin{figure}[t!]
\begin{center}
\begin{minipage}{.6\textwidth}
\includegraphics[width=\textwidth]{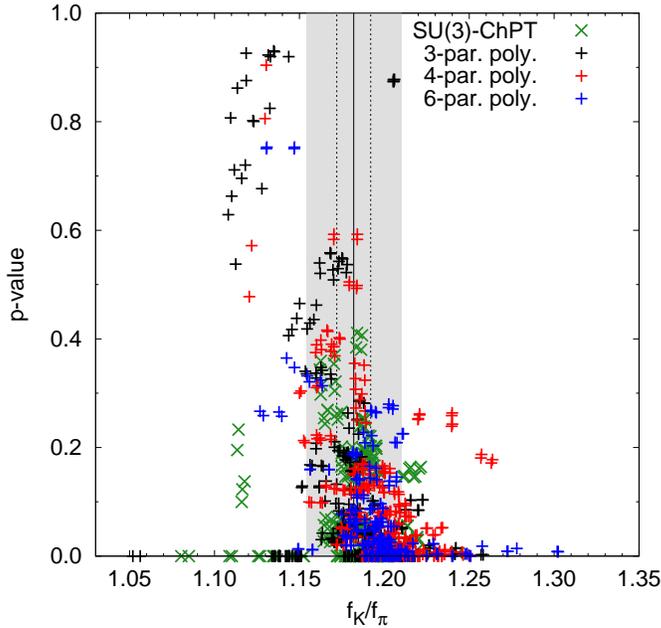}
\end{minipage}%
\hspace{.05\textwidth}
\begin{minipage}{.3\textwidth}
\caption{Scatter plot of the $p$-value obtained in each of the 1368 performed fits against the extrapolated (physical masses, infinite-volume and continuum limit) decay-constant ratio $f_K/f_\pi$ from that particular fit. The {\it solid black line, dashed black lines, and the grey-shaded area} indicate the central value, statistical and combined error, resp., of our final result, Eq.~(\protect\ref{eq:fkfPi_final}).}
\label{fig:scatter}
\end{minipage}
\end{center}
\end{figure}

In Fig.~\ref{fig:scatter} we show the $p$-values obtained in different fits, which were discussed in Sec.~\ref{subsec:fits}, plotted against the decay-constant ratio at the physical point (physical masses, infinite-volume limit, continuum limit) as obtained from a specific fit. In order to quantify the systematic uncertainty from our extrapolation to the physical point, we look both at the $p$-value weighted average and its variance and at an unweighted (``flat'') average, finding
\begin{equation}
\left.\frac{f_K}{f_\pi}\right|_{p{\rm -value}} \;=\; 1.173(11)_{\rm stat}(29)_{\rm syst}\,,\;\;\;\left.\frac{f_K}{f_\pi}\right|_{\rm flat} \;=\; 1.191(08)_{\rm stat}(24)_{\rm syst}\,.
\end{equation}
In order to try to reduce any bias introduced by our choice our fitting procedures etc. for our final value of the decay-constant ratio we take the straight average of these two resulting in
\begin{equation}
\label{eq:fkfPi_final}
\frac{f_K}{f_\pi} \;=\; 1.182(10)_{\rm stat}(26)_{\rm syst}\;=\;1.182(28)_{\rm comb}\,.
\end{equation}
The implications of this result for the CKM-matrix elements will be discussed in Sec.~\ref{sec:disc}.

\begin{figure}[t!]
\begin{center}
\includegraphics[width=.45\textwidth]{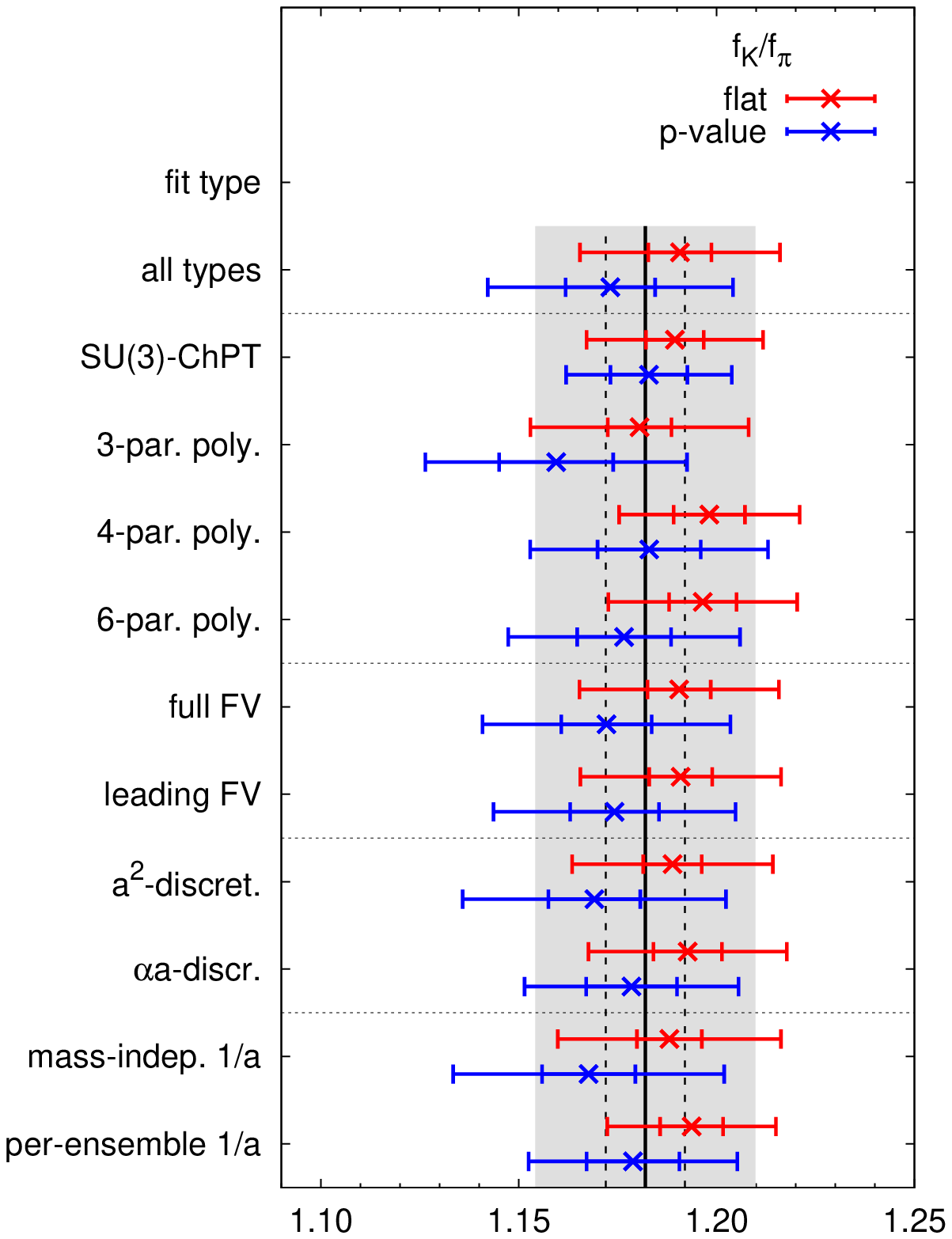}%
\hspace{.05\textwidth}%
\includegraphics[width=.45\textwidth]{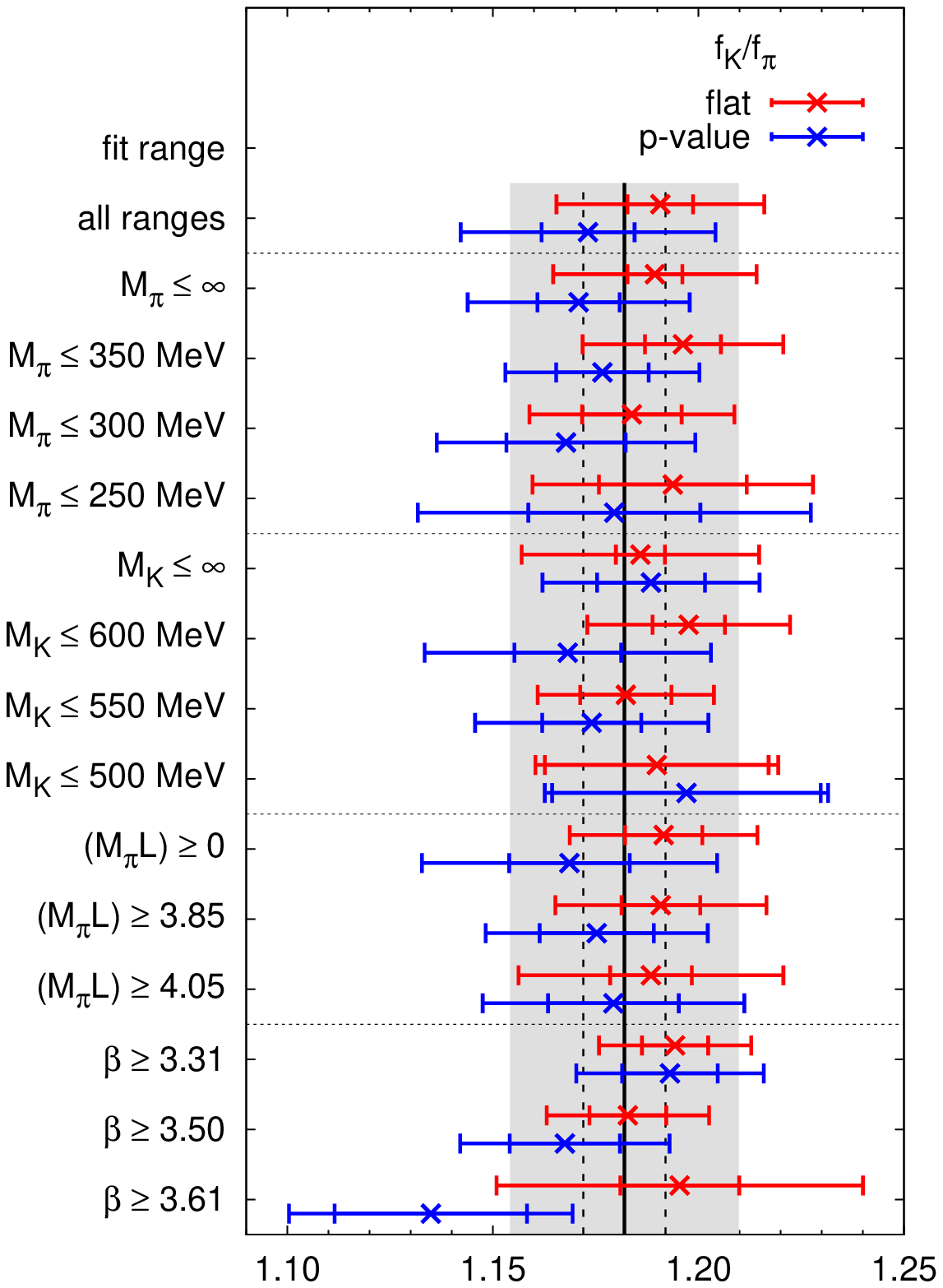}
\cCap
\caption{Results obtained without weighting {\it (red symbols)} and with $p$-value weighting {\it (blue symbols)} from various subsets of our fits. The {\it left panel} shows subsets given by the choices of fit forms, while the {\it right panel} shows subsets given by different fit ranges. The {\it solid black line, dashed black lines, and the grey-shaded area} indicate the central value, statistical and combined error, resp., of our final result, Eq.~(\protect\ref{eq:fkfPi_final}).}
\label{fig:fit_types_ranges}
\end{center}
\end{figure}

We also looked at different subsets of our fits, sorted either by the fit form used for the mass interpolation, the continuum-limit extrapolation and so on or sorted by the fit ranges in $M_\pi$, $M_K$ etc. The results are shown in the two panels of Fig.~\ref{fig:fit_types_ranges} (a detailed table of these results can be found in \cite{Durr:2016ulb}). While most subsets in each group show comparable results, the biggest variation seems to stem from the variation in the cut to the lattice spacing $a$ equivalent to the gauge coupling $\beta$ (group at the bottom of the right panel of Fig.~\ref{fig:fit_types_ranges}).

\aFig

\aFig

\section{Discussion}
\label{sec:disc}
\cSec

Since in Eq.~(\ref{eq:marciano}) the real-world ratio $f_{K^\pm}/f_{\pi^\pm}$ appears, which also includes electromagnetic and $m_{\rm up}\neq m_{\rm down}$ effects, we have to correct our result for $f_K/f_\pi$, Eq.~(\ref{eq:fkfPi_final}), to account for these effects. An analysis \cite{Gasser:1984gg,Cirigliano:2011tm} utilizing ChPT shows, that this correction can be obtained in the following way
\begin{equation}
\label{eq:isoCorrection}
\frac{f_{K^\pm}}{f_{\pi^\pm}}\;=\;\frac{f_K}{f_\pi}\,\sqrt{1\,+\,\delta_{\rm SU(2)}}\,,
\end{equation}
where the correction term was estimated within ChPT \cite{Cirigliano:2011tm} to have the value $\delta_{\rm SU(2)} =\linebreak[4] -0.0043(12)$. The FLAG-report \cite{Aoki:2013ldr}\footnote{The same holds true for the values reported in the latest FLAG report \cite{Aoki:2016frl} which appeared recently as a preprint.} quotes very similar values from lattice simulations with $N_f=2+1$ flavors. Also, this correction was determined for $N_f=2$ flavors in \cite{deDivitiis:2011eh} to be $\delta_{\rm SU(2)} = -0.0078(7)$. Conservatively, we decided to use the average of these two values with a 100 per cent uncertainty: $\delta_{\rm SU(2)}=-0.0061(61)$ in Eq.~(\ref{eq:isoCorrection}) to correct our result, Eq.~(\ref{eq:fkfPi_final}). We obtain
\begin{equation}
\label{eq:fkfPi_isoCorr}
\frac{f_{K^\pm}}{f_{\pi^\pm}}\;=\; 1.178(10)_{\rm stat}(26)_{\rm syst}\;=\;1.178(28)_{\rm comb}\,.
\end{equation}

Now, we are able to use our value, Eq.~(\ref{eq:fkfPi_isoCorr}), together with the experimental result \cite{Moulson:2014cra,Rosner:2015wva}
\begin{equation}
\frac{V_{\rm us}}{V_{\rm ud}} \, \frac{f_{K^\pm}}{f_{\pi^\pm}}\;=\;0.27599(29)(24)\;=\;0.27599(38)_{\rm exp}
\end{equation}
to determine the ratio of CKM-matrix elements
\begin{equation}
\label{eq:Vratio}
\frac{V_{\rm us}}{V_{\rm ud}} \;=\; 0.2343(20)_{\rm stat}(52)_{\rm syst}(03)_{\rm exp}\;=\;0.2343(55)_{\rm comb}\,.
\end{equation}

If one assumes first-row unitarity $V_{\rm ud}^2+V_{\rm us}^2+|V_{\rm ub}|^2=1$ of the CKM-matrix together with the result from \cite{Rosner:2015wva} that $|V_{\rm ub}|=4.12(37)(06)\,\cdot\,10^{-3}$  is very small compared to the other two matrix elements, we obtain from our result, Eq.~(\ref{eq:Vratio}), 
\begin{equation}
\label{eq:VudVus}
V_{\rm ud}\;=\;0.9736(04)_{\rm stat}(11)_{\rm syst}(01)_{\rm exp}\,,\;\;\;V_{\rm us}\;=\;0.2281(18)_{\rm stat}(48)_{\rm syst}(03)_{\rm exp}\,.
\end{equation}
Alternatively, one may use the high-precision result from super-allowed nuclear $\beta$-decays \cite{Hardy:2014qxa} $V_{\rm ud} = 0.97417(21)_{\rm nuc}$ to obtain from Eq.~(\ref{eq:Vratio})
\begin{equation}
V_{\rm us} \;=\; 0.2282(19)_{\rm stat}(51)_{\rm syst}(03)_{\rm exp+nuc}\,.
\end{equation}
and use these two results together with the above $|V_{\rm ub}|$ to check for first-row unitarity
\begin{equation}
V_{\rm ud}^2+V_{\rm us}^2+|V_{\rm ub}|^2\,-\,1\;=\;0.0011(09)_{\rm stat}(23)_{\rm syst}(05)_{\rm exp+nuc}\;=\;0.0011(25)_{\rm comb}\,,
\end{equation}
showing that within the errors the assumption of first-row unitarity of the CKM-matrix is fulfilled.

Comparing our result, Eq.~(\ref{eq:fkfPi_isoCorr}), with the average from the latest FLAG-report \cite{Aoki:2016frl} for $N_f=2+1$ lattice-QCD simulations: $f_{K^\pm}/f_{\pi^\pm}=1.192(5)$, one observes that the two agree within our large uncertainty, although our result favors a smaller decay-constant ratio. Definitely, our uncertainty needs to be reduced for one to be able to make a more robust statement. The same holds true if one compares our results for the CKM-matrix elements, Eq.~(\ref{eq:VudVus}), with the latest $N_f=2+1$ FLAG-averages \cite{Aoki:2016frl}: $V_{\rm ud}=0.97451(23)$, $V_{\rm us}=0.2243(10)$. For that reason, currently we are examining in which way different choices of fit ranges, fit forms might help to improve the precision of our result.

\vspace*{.3cm}
\noindent{\bf Acknowledgment} We thank our colleagues Zolt\'an Fodor, Christian H\"olbling, Stefan Krieg,\linebreak[4] Laurent Lellouch, Thomas Lippert, Thomas Rae, Andreas Sch\"afer, K\'alm\'an K.~Szab\'o, and Lukas Varnhost for the collaboration on this project.

Computations were carried out on the BG/Q supercomputer JUQUEEN at Forschungszentrum J\"ulich through a NIC grant, on Turing at IDRIS, France, under GENCI-IDRIS grant 52275, and on a local cluster at the University of Wuppertal.
This work was supported by the DFG through the SFB/TRR 55 ``Hadron Physics from Lattice QCD'', by the EU Framework Programme 7 grant (FP7/2007-2013)/ERC No 208740, by the OTKA grant OTKA-NF-104034, and by the projects OCEVU Labex (ANR-11-LABX-0060) and A*MIDEX (ANR-11-IDEX-0001-02).
In addition, E.E.S.\ acknowledges support from the EU grant PIRG07-GA-2010-268367.



\bibliography{references}
\bibliographystyle{h-physrev5}

\end{document}